\documentclass[runningheads]{llncs}
\usepackage{graphicx}
\usepackage{xcolor}
\usepackage{hyperref}
 \usepackage{multirow}
 \usepackage{amsmath}
  \usepackage{wrapfig,lipsum,booktabs}
\begin{document}
\title{Enhance Topics Analysis based on Keywords Properties}
%
%
%
%


\author{Antonio Penta\\
%
%
\institute{Analytics \& AI Group, R\&D Global Innovation Center -The Dock Accenture}
\email{antonio.penta@accenture.com}}

\maketitle              
\begin{abstract}
Topic Modelling is one of the most prevalent text analysis technique used to explore and retrieve collection of documents.
The evaluation of the topic model algorithms is still a very challenging tasks due to the absence of gold-standard list of topics to compare against for every corpus. In this work, we present a specificity score based on keywords properties that is able to select the most informative topics. This approach helps the user to focus on the most informative topics. In the experiments, we show that we are able to compress the state-of-the-art topic modelling results of different factors with an information loss that is much lower than the solution based on the recent coherence score presented in literature.
\end{abstract}

\section{Introduction}

Topic modelling assumes that we can describe documents using a mixture of topics, where each topic is represented by a probability distribution over the corpus vocabulary, and each document can be represented by multiple topics shared across the corpus. The advantage of this approach is that it does not require any specific upfront information.  Currently, topic modelling is the most prevalent text analysis technique used to explore and retrieve collection of documents, and we have seen its application in many areas: COVID-19 pandemic~\cite{Doanvo2020}, marketing~\cite{Reisenbichler2018}, bionInformatics~\cite{Liu2016}, along with many other applications~\cite{BoydGraber2017}. The evaluation of the topic model algorithms has been always very challenging, mostly because there is no gold-standard list of topics to compare against for every corpus. Thus, evaluating the latent space of topic models requires defining new measures. For example it is often unclear, whether an observed difference is due to one model being better than another. In our work we address the problem of filtering the topics results provided by the state-of-the art topic modelling (LDA~\cite{Blei03}), focusing the user attention on the ones with better quality. In Table~\ref{tab:sota}, we report the current approaches to measure topic quality categorised by evaluation type. Our work extends the current literature presenting a score named \emph{specificity} based on keywords properties. In the next subsections, we details the rational behind these scores, and then we present some experiments.

\begin{table}[]
\centering
\begin{tabular}{|l|l|l|l|}
\hline
\multicolumn{1}{|c|}{\textbf{Types}} &
  \multicolumn{1}{c|}{\textbf{Approaches}} &
  \multicolumn{1}{c|}{\textbf{Methods}} &
  \multicolumn{1}{c|}{\textbf{References}} \\ \hline
\multirow{10}{*}{\begin{tabular}[c]{@{}l@{}}Intrinsic \\ Evaluation\end{tabular}} &
  \multirow{4}{*}{\begin{tabular}[c]{@{}l@{}}Proximity\\ Measure\end{tabular}} &
\begin{tabular}[c]{@{}l@{}}Language Modelling \\ (Perplexity,Document Completion)\end{tabular}&\cite{Blei03}, \cite{Wallach2009} \\ \cline{3-4} 
 &
   &
 \begin{tabular}[c]{@{}l@{}}Topic Coherence with Scores \\ (UCI-UMASS-NGD)\end{tabular}&
  \cite{mimno-2011}, \cite{Newman2010}, \cite{stevens-etal-2012},\cite{roder2015} \\ \cline{3-4} 
 &
   &
  Topic Coherence with Embeddings &
\cite{ding-etal-2018-coherence} \\ \cline{3-4} 
 &
   &
  Topic Uniqueness, Diversity &
\cite{topicuni}, \cite{Dieng} \\ \cline{2-4} 
 &
  \multirow{3}{*}{\begin{tabular}[c]{@{}l@{}}Human-in-the-loop \\ /Crowdsourcing\end{tabular}} &
  Word Intrusion &
\cite{Chang2009} \\ \cline{3-4} 
 &
   &
  Questionnaire &\cite{Chang2009} \\ \cline{3-4} 
 &
   &
  Topic  Consensus &\cite{Morstatter} \\  \cline{2-4}
  &
  \multirow{1}{*}{Diagnostic} &
  Topic Size, Word Length  &
\cite{Boyd-Graber:Mimno:Newman-2014} \\ \cline{3-4} 
  \hline
\multirow{3}{*}{\begin{tabular}[c]{@{}l@{}}Extrinsic \\ Evaluation\end{tabular}} &
  \multirow{3}{*}{\begin{tabular}[c]{@{}l@{}}Downstream\\Tasks\end{tabular}} &
  Document Classification &
\cite{nguyen-etal-2015} \\ \cline{3-4} 
 &
   &
  Document Clustering &
\cite{nguyen-etal-2015} \\ \cline{3-4} 
 &
   &
  Topic Ranking &
\cite{bhatia-etal-2017} \\ \hline
\end{tabular}
\caption{Classification of the approaches used to evaluate Topic Models along with the relevant state-of-the-art related works.}
\label{tab:sota}
\end{table}

\vspace{-0.5cm}
\section{Topic Filtering}\label{sec:filtering}

The goal of the Topic Filtering is to select topics based on the value of their content, and 
we assume that the content of each topic is described by its Top-K words. The value of the words can be measured by looking at different properties, for example semantic, relevance, representativeness, centrality, rarity and so on. However, all those properties require proper definitions and related proxy measures such as frequency in a corpus and/or structure properties of the associated node in a knowledge base. In the next subsections, we detailed the properties used in this work. 


\subsection{Relevant Rare and Relevant Frequent Words}
Relevance is one of the fundamental concepts in Information Retrieval, and it can be defined in different ways~\cite{Cooper1971},\cite{Robertson2009}. For example, a relevance of an object has been measured mostly in relations of the value of its information. In our case, the object are the words, the information are their semantics and the value is the domain knowledge that a user will gain by observing a new word respect to an initial set. In this regard, measuring the relevance of a word requires capturing its knowledge acquisition power.  In order to do that, we reframe the knowledge acquisition feature of the words in the terms of the following  words characteristics: \emph{relevant rarity} and \emph{relevant frequent}. A word with a high relevant rarity score is a word used to represent the particularities of a specialised domain versus a common one. For example, if we are interested in retrieving specialised knowledge from a collection of conversations among customers and agents,  we would like to extract the words that are related to problems/products/codes rather than names of the customers. Both words can be rare (in terms of frequency of usage) but the former are unique of the domain of interest, therefore they are relevant, while the latter can also be unique (i.e. names of persons), but they are less useful to learn more about the business domain, therefore less relevant.  In our case, we use the Residual Inverse Document Frequency~(R-IDF)(~\cite{Church1999}) to measure the relevance rare property. .We are not only interested in rare relevant words that can provide insights of the domain, but also in the words that are driving the domain discussion. These words can be frequent in the data collection because their usage is related to how the user structures the information. For example, if we have a collection of conversational data for supporting a troubleshooting inquiry to a call centre support desk, the word  ``problem" can appear quite frequent and it could be judged as non-important using IDF-based approach, but it is relevant for understanding the context in which a product code (relevant rare word) is mentioned in the conversation, therefore topics that contain the ``problem"  word could be useful for the  comprehension of the domain. For extracting words with the relevant frequent property, we use the support measure used in the association rule theory~(\cite{Tan}). In particular we are interested in finding the words that are frequent enough in the collection and that are relevant in a context, which means that their presence is constraining the frequency of the next words. We measure this property using the 
following score \emph{Prevalent Association Lift-(PAL)}:
\begin{equation}
PAL(w_l,w_r,\mathcal{C}_n)=\frac{\overrightarrow{S}(w_l,w_r,\mathcal{C}_n)}{\sqrt{S(w_l,\mathcal{C}_n)+S(w_r,\mathcal{C}_n)}}
\end{equation}
where $w_l$, $w_r$ are two words, named \emph{left} and \emph{right} words, $\mathcal{C}_n$ is a collection of sequences of $n$ words extracted from the text, which is named \emph{context}, $S(w,\mathcal{C}_n)$ is the support of the word $w$ in the context $\mathcal{C}_n$, and $\overrightarrow{S}(w_r,w_l,\mathcal{C}_n)$ is the support of the rule $w_r\rightarrow w_l$ in the context $\mathcal{C}_n$. We note that the above scores is higher in the presence of associations frequent $\rightarrow$ rare words, rather than  frequent $\rightarrow$ frequent words. From the score above, we can measure the relevant frequent property of a word $w$ in the context $\mathcal{C}$ using the max value of the PAL: $max_{w^*\in(\mathcal{C})}(PAL(w,w^*,\mathcal{C}))$.

\subsection{Coherence as Proximity Dependency}
Topics are mostly presented to the users as a set of words, or a ranked list of words. The users are connect these words using an internal model of the world, which depends from their personal knowledge and can vary in each subject. However, in most of the use-cases the purpose of topic analysis is to discover and summarise only the knowledge stored in the text collection, which could be unknown or partially known from the user. Therefore, the objective of a Topic Filter is to show the knowledge in the data by selecting the topics whose content facilitates the user reasoning process (what we called interaction), making it less dependent on some pre-knowledge, avoiding the consumption of the topics analysis only from domain experts. Most of the time, user reasoning process has the role of filling the semantic gap across words, connect words with missing knowledge and abstract the content using the words as hints. In order to facilitate this process, we are interested to highlight topics where is easier to interconnect the words without assuming pre-existing expert knowledge.  The frequency-based coherence measures used in literature are done at document level for computational reasoning; therefore, they do not properly weigh words that are in closer proximity, so for this reason we define the \emph{coherence}~($Coh$) among the Top-K words in a topic as follows:
\begin{equation}
\begin{split}
Coh(w_1,\ldots,w_K)=\frac{\sum_{i=1}^{K}\sum_{j=1,i\neq j}^{K}\alpha CosSim(\phi(w_i),\phi(w_j))}{2C^k_2} \\
\end{split}
\end{equation}
where $\alpha$=$max(R\textrm{-}IDF(w_i),R\textrm{-}IDF(w_j))$, $\phi$ is function that retrieves the embeddings for the input words , $CosSim$ is the cosine similarity function between two arrays, $C^k_2$ is the binomial coefficient. The R-IDF
is used to give more weight to rare terms, because common terms due to their nature (they appear frequently in the data) have embeddings that are kind of centroids, therefore they can bias the coherence score toward topics that contain more common words, which are less important in understanding the domain knowledge.

\subsection{Specificity}
We combine the above word properties/scores to have an unique score for each topic, that we named \emph{Specificity}~($Sp$), and it is defined as follows:
\begin{equation}
Sp(w_1,\ldots,w_K)=Coh(w_1,\ldots,w_K)*\sum_{w\in\{w_1,\ldots,w_K\}}PAL(w)
\end{equation}
 where the $\{w_1,\ldots,w_K\}$ is the set of Top-K words selected for each topic.

\section{Experiments}
For the experiments, we use a version of the Review Amazon dataset\footnote{https://nijianmo.github.io/amazon/index.html} that does not contain categories with more than 1M reviews, to have a comparable set of textual data. For each category, we also select the first 1000 reviews with larger text. The text collection (12 categories, 1000 samples for each category) is pre-processed using a standard NLP approaches( i.e. lemmatization, standard stop words), we also remove adjectives and adverbs to have conceptual topics. We run LDA\footnote{https://radimrehurek.com/gensim/models/ldamodel.html}, fixing the number of iterations and passes equal to the 20\% of the length of the whole collection. These parameters help us to have low perplexity models.  For the number of topics, we select K=\{L,2L,5L,10L,15L\}, where L is the number of categories (i.e. labels) in the dataset. For each K, we evaluate the F1 accuracy using a Random Forest~(RF) classifier trained to predict the category for each review. We use the topic distribution as a feature. For the RF, we use the default options, 100 trees, and 0.5\% data as a training set. The F1 accuracy is reported using k-fold cross-validation (k=5). For each topic, we use the first 25 words to compute the properties /score. For PAL, we use a context of 5 words, and for the word-embedding, we use the Skip-gram approach because the proximity condition is better represented. 
\subsection{Methodology and Results}\label{sec:results}
The objective is to evaluate if the selected topics are providing enough information to classify correctly the category. In particular, we use the following indicator to measure the quality of our filtering strategies: DataCompression (DC)=$\frac{|K|}{|K^*|}$, where $K$ is the full number of topics,  and $K^*$ is the number of topics selected using one of the approaches. 
The idea is to measure what is the Information Loss in the classification process for different data compression values. In other words, if I select only the filtered topics for training a classifier what will be the loss in the classification accuracy respect the case where we consider all the topics. To measure that, we define the Information Loss as follows:$1-\frac{F1}{F1^*}$, where $F1^*$ is the F1 measure of a reference model, while $F1$ is the F1 accraucy obtained using the compressed model~(i.e. with only the selected topics as features).
\begin{wraptable}{r}{0.5\textwidth}
\centering
\begin{tabular}{|l|l|l|l|}
\hline
\multicolumn{1}{|c|}{LM} & \multicolumn{1}{c|}{K} & \multicolumn{1}{c|}{Pr} & \multicolumn{1}{c|}{F1}    \\ \hline
L   & 12  & -7.554 & 0.780 (+/- 0.006) \\ \hline
\textbf{~2L}   & \textbf{40}            & \textbf{-7.540}         & \textbf{0.830 (+/- 0.005)} \\ \hline
5L  & 60  & -7.522 & 0.846 (+/- 0.006) \\ \hline
10L & 120 & -7.456 & 0.851 (+/- 0.007) \\ \hline
15L & 180 & -7.489 & 0.844 (+/- 0.006) \\ \hline
\end{tabular}
\caption{LDA results for the classification task for different Number of topics (K), F1 is the F1-Accuracy, Pr is the perplexity. LM specifies K as number of categories (i.e labels) in the dataset.}
\label{tab:f1}
\end{wraptable} 
In Table~\ref{tab:f1}, we report all the LDA models with different values of K along with their perplexity values and F1 accuracies. The results show that increasing the number of topics does not improve the accuracy of the model. For example, the model with K=10L has the same F1 accuracy as the model with K=2L, which means that the added topics do not provide useful information to discriminate the categories. We compute the Information Loss of the model with K=5L,10L, and 15 using the model with K=2L as a reference model.  For example, we compute the Information Loss of the model with K=5L in the case we select only 40 topics as input for the classier having a data compression factor of the model equal to  2.5. In Table~\ref{tab:results}, we report all the Information Losses (IL)s for each approach and model. IL-Rnd refers to the approach based on the random selection of the topics, IL-R-IDF, IL-PAL, IL-Coh, IL-Sp refers to the approaches based on the R-IDF, PAL, Coherence, Specificity values respectively.  We also include results using the coherence formulation defined in ~\cite{ding-etal-2018-coherence}, and also the one based on the geometric means of the R-IDF, PAL, Coh values for each topic. We note that our approach based specificity outperforms the baselines (random and \cite{ding-etal-2018-coherence} for different data compression values.

\begin{table}[h]
\begin{tabular}{|c|l|l|l|l|l|l|l|l|}
\hline
Model &
  \multicolumn{1}{c|}{DC} &
  \multicolumn{1}{c|}{IL-Rnd} &
  \multicolumn{1}{c|}{\textbf{IL-\cite{ding-etal-2018-coherence}}} &
  \multicolumn{1}{c|}{IL-RIDF} &
  \multicolumn{1}{c|}{IL-PAL} &
  \multicolumn{1}{c|}{IL-Coh} &
  \multicolumn{1}{c|}{IL-GM} &
  \multicolumn{1}{c|}{\textbf{IL-Sp}} \\ \hline
\begin{tabular}[c]{@{}c@{}}5L\\ ($K$=60, $K^*$=40)\end{tabular}   & 2.5 & 13,821\% & \textbf{0\%}      & 0\%      & 0\%      & 0\%      & 0\%      & \textbf{0\%}      \\ \hline
\begin{tabular}[c]{@{}c@{}}5L \\($K$=60, $K^*$=10)\end{tabular}   & 6   & 60,721\% & \textbf{43,885\%} & 3,116\%  & 41,445\% & 39,713   & 34,857\% & \textbf{35,848\%} \\ \hline\hline
\begin{tabular}[c]{@{}c@{}}10L\\($K$=120,$K^*$=40)\end{tabular} & 3   & 45,60\%  & \textbf{29,642\%} & 6,21\%   & 9,854\%  & 8,87\%   & 6,280\%  & \textbf{8,441\%}  \\ \hline
\begin{tabular}[c]{@{}c@{}}10L\\($K$=120,$K^*$=10)\end{tabular}  & 12  & 79,321\% & \textbf{73,448\%} & 59,614\% & 62,093\% & 59,401\% & 55,069   & \textbf{52,936\%} \\ \hline\hline
\begin{tabular}[c]{@{}c@{}}15L\\($K$=180,$K^*$=40)\end{tabular} & 4.5 & 50,362\% & \textbf{29,661\%} & 6,165\%  & 9,687\%  & 8,927\%  & 6,183\%  & \textbf{8,413\%}  \\ \hline
\begin{tabular}[c]{@{}c@{}}15L\\($K$=180,$K^*$=10)\end{tabular} & 18  & 79,136\% & \textbf{73,461\%} & 59,760\% & 61,937\% & 59,456\% & 55,015\% & \textbf{53,015\%} \\ \hline
\end{tabular}
\caption{The Table describes the results of the different approaches proposed in the paper presented as Informal Loss (IL) respect to the reference model  with K=2L. Lower is value of the Information Loss better is the selection strategy.  $K^*$ refers to the number of topics selected for the compressed model. The attributes of the tables are described in Section~\ref{sec:results}.}
\label{tab:results}
\end{table}

\section{Conclusion}\label{sec:conclusion}
In this work, we have presented a new approach for evaluating the quality of the topics based on the specificity of the Top-K words belonging to each topic. We discussed the rationale behind this measure, and we provided some experiments that showed how the selected topics based on the specificity have an information loss that is inferior to the approaches based on coherence scores presented in the recent literature.
\bibliographystyle{splncs04}
\bibliography{paper}
\end{document}